\documentclass{article}
\usepackage{cite}
\usepackage{graphicx}
\usepackage{dcolumn}

\begin{document}

\date{}
\title{Algebraic treatment of the Bateman Hamiltonian}
\author{Francisco M. Fern\'{a}ndez \thanks{%
E-mail: fernande@quimica.unlp.edu.ar} \\
INIFTA, Divisi\'on Qu\'imica Te\'orica,\\
Blvd. 113 S/N, Sucursal 4, Casilla de Correo 16,\\
1900 La Plata, Argentina.}
\maketitle

\begin{abstract}
We apply the algebraic method to the Bateman Hamiltonian and obtain its
natural frequencies and ladder operators from the adjoint or regular matrix
representation of that operator. Present analysis shows that the
eigenfunctions compatible with the complex eigenvalues obtained earlier by
other authors are not square integrable. In addition to this, the ladder
operators annihilate an infinite number of such eigenfunctions.
\end{abstract}

\section{Introduction}

\label{sec:intro}

There has recently been a controversy about the eigenvectors and eigenvalues
of the Bateman Hamiltonian\cite{B31,FT77}. On one side, Deguchi et al\cite
{DFN19,DF19} discussed two quantization approaches for the Bateman
Hamiltonian based on ladder operators, on the other, Bagarello et al\cite
{BGR19,BGR20} argued that there is no square integrable vacuum for the
natural ladder operators.

The purpose of this paper is the study of the problem by means of the
algebraic method\cite{FC96} that has proved suitable for the treatment of a
variety of quadratic Hamiltonian operators\cite{F15a,F15b,F16a,F16b,F16c,F19}%
.

In section~\ref{sec:model} we briefly discuss the model and derive a simpler
dimensionless version of the Hamiltonian operator. In section~\ref
{sec:algebraic} we outline the main ideas about the algebraic method.
Finally, in section~\ref{sec:Results_discussion} we apply this approach to
the Bateman Hamiltonian, show results and draw conclusions.

\section{The model}

\label{sec:model}

The Hamiltonian operator for the so called Bateman oscillator model is\cite
{DFN19,DF19,BGR19,BGR20}
\begin{equation}
H=\frac{1}{2m}\left( p_{1}^{2}-p_{2}^{2}\right) +\frac{m\omega ^{2}}{2}%
\left( x_{1}^{2}-x_{2}^{2}\right) -\frac{\gamma }{2m}\left(
x_{1}p_{2}+x_{2}p_{1}\right) ,  \label{eq:H_Bate}
\end{equation}
where $m,\gamma ,\omega ^{2}>0$ and $\left[ x_{j},p_{k}\right] =i\hbar
\delta _{jk}$.

It is convenient to transform the Hamiltonian operator (\ref{eq:H_Bate})
into a dimensionless one. To this end we define the new variables $\left(
x_{1},x_{2},p_{1},p_{2}\right) \rightarrow \left( \alpha x,\alpha y,\hbar
\alpha ^{-1}p_{x},\hbar \alpha ^{-1}p_{y}\right) $, so that $\left[
u,p_{v}\right] =i\delta _{uv}$, $u,v=x,y$. On choosing $\alpha ^{2}=\hbar
/(m\omega )$ we are left with the one-parameter Hamiltonian
\begin{equation}
H_{d}=\frac{H}{\hbar \omega }=\frac{1}{2}\left( p_{x}^{2}-p_{y}^{2}\right) +%
\frac{1}{2}\left( x^{2}-y^{2}\right) -\frac{b}{2}\left( xp_{y}+yp_{x}\right)
,  \label{eq:H_dim}
\end{equation}
where $b=\gamma /(m\omega )$. It is worth noticing that $%
[H_{d},H_{0}]=[H_{d},H_{1}]=[H_{0},H_{1}]=0$, where $H_{0}=H_{d}(b=0)$ and $%
H_{1}=H-H_{0}$.

The operator (\ref{eq:H_dim}) satisfies
\begin{equation}
\left\langle f\right| H_{d}\left| g\right\rangle =\int_{-\infty }^{\infty
}\int_{-\infty }^{\infty }f(x,y)^{*}H_{d}g(x,y)dxdy=\int_{-\infty }^{\infty
}\int_{-\infty }^{\infty }\left[ H_{d}f(x,y)\right]
^{*}g(x,y)dxdy=\left\langle H_{d}f\right| \left. g\right\rangle ,
\label{eq:Hermit_inner_prod}
\end{equation}
for every pair of square-integrable functions $f(x,y)$ and $g(x,y)$.
Therefore, from now on we use the standard quantum-mechanical Hermitian
notation $H_{d}^{\dagger }=H_{d}$.

\section{The algebraic method}

\label{sec:algebraic}

In this section we outline the algebraic method discussed in previous papers%
\cite{F15a,F15b,F16a,F16b,F16c,F19}. The model discussed in section~\ref
{sec:model} is a particular case of a quadratic Hamiltonian of the form
\begin{equation}
H=\sum_{i=1}^{2K}\sum_{j=1}^{2K}\gamma _{ij}O_{i}O_{j},
\label{eq:H_quadratic}
\end{equation}
where $\left\{ O_{1},O_{2},\ldots ,O_{2K}\right\} =\left\{
x_{1},x_{2},\ldots ,x_{K},p_{1},p_{2},\ldots ,p_{K}\right\} $, $%
[x_{m},p_{n}]=i\delta _{mn}$, and $[x_{m},x_{n}]=[p_{m},p_{n}]=0$. Here we
restrict ourselves to operators that satisfy $H^{\dagger }=H$ but the
approach applies also to other cases as well\cite{F16c}. The algebraic
method is particularly useful for the analysis of the spectrum of $H$
because it satisfies the commutation relations
\begin{equation}
\lbrack H,O_{i}]=\sum_{j=1}^{2K}H_{ji}O_{j}.  \label{eq:[H,Oi]}
\end{equation}
For this reason it is possible to obtain an operator of the form
\begin{equation}
Z=\sum_{i=1}^{2K}c_{i}O_{i},  \label{eq:Z}
\end{equation}
such that
\begin{equation}
\lbrack H,Z]=\lambda Z.  \label{eq:[H,Z]}
\end{equation}
The operator $Z$ is important for our purposes because if $H\left| \psi
\right\rangle =E\left| \psi \right\rangle $ then
\begin{equation}
HZ\left| \psi \right\rangle =(E+\lambda )Z\left| \psi \right\rangle .
\label{eq:HZ|Psi>}
\end{equation}

It follows from equations (\ref{eq:[H,Oi]}), (\ref{eq:Z}) and (\ref{eq:[H,Z]}%
) that
\begin{equation}
(\mathbf{H}-\lambda \mathbf{I})\mathbf{C}=0,  \label{eq:HC=-lambdaC}
\end{equation}
where $\mathbf{H}$ is a $2K\times 2K$ matrix with elements $H_{ij}$, $%
\mathbf{I}$ is the $2K\times 2K$ identity matrix and $\mathbf{C}$ is a $%
2K\times 1$ column matrix with elements $c_{i}$. There are nontrivial
solutions only for those values of $\lambda $ that are roots of the
characteristic polynomial $p(\lambda )=\det (\mathbf{H}-\lambda \mathbf{I})$%
. $\mathbf{H}$ is called the adjoint or regular matrix representation of $H$
in the operator basis set $\{O_{1},O_{2},\ldots ,O_{2K}\}$\cite{FC96}. This
matrix is closely related to the fundamental matrix that proved to be useful
in determining the conditions under which a PT-symmetric elliptic quadratic
differential operator with real spectrum is similar to a self-adjoint
operator\cite{CGHS12}.

In the case of an Hermitian operator we expect all the roots $\lambda _{i}$,
$i=1,2,\ldots ,2K$ of the characteristic polynomial $p(\lambda )$ to be
real. These roots are obviously the natural frequencies of the
quantum-mechanical system (the actual quantum-mechanical frequencies being
linear combinations of them). However, in the present case we do not assume
this condition. Therefore, it follows from equation (\ref{eq:[H,Z]}) that
\begin{equation}
\lbrack H,Z^{\dagger }]=-\lambda ^{*}Z^{\dagger },  \label{eq:[H,Z+]}
\end{equation}
where $Z^{\dagger }$ is a linear combination like (\ref{eq:Z}) with
coefficients $c_{i}^{*}$. This equation tells us that if $\lambda $ is a
real root of $p(\lambda )=0$, then $-\lambda ^{*}$ is also a root.
Obviously, $Z$ and $Z^{\dagger }$ look like a pair of annihilation-creation
or ladder operators because, in addition to (\ref{eq:HZ|Psi>}), we also have
\begin{equation}
HZ^{\dagger }\left| \psi \right\rangle =(E-\lambda ^{*})Z^{\dagger }\left|
\psi \right\rangle .  \label{eq:HZ+|Psi>}
\end{equation}

If $A$ is a quadratic Hermitian operator that satisfies
\begin{eqnarray}
\left[ H,A\right] &=&0,  \nonumber \\
\left[ A,O_{i}\right] &=&\sum_{j=1}^{2K}A_{ji}O_{j},  \label{eq:[H,A]=0}
\end{eqnarray}
then it follows from the Jacobi identity
\begin{equation}
\left[ H,\left[ A,O_{i}\right] \right] +\left[ O_{i},\left[ H,A\right]
\right] +\left[ A,\left[ O_{i},H\right] \right] ,
\end{equation}
that
\begin{equation}
\mathbf{HA}-\mathbf{AH}=\mathbf{0},  \label{eq:[mH,mA]=0}
\end{equation}
where $\mathbf{A}$ is the adjoint matrix representation of $A$.

\section{Results and discussion}

\label{sec:Results_discussion}

The eigenvalues of the matrix representation $\mathbf{H}$ of $H_{d}$
\begin{equation}
\mathbf{H}=\frac{i}{2}\left(
\begin{array}{llll}
0 & b & 2 & 0 \\
b & 0 & 0 & -2 \\
-2 & 0 & 0 & -b \\
0 & 2 & -b & 0
\end{array}
\right) ,
\end{equation}
are
\begin{equation}
\lambda _{1}=-1-i\frac{b}{2},\;\lambda _{2}=-1+i\frac{b}{2},\;\lambda
_{3}=1-i\frac{b}{2},\;\lambda _{4}=1+i\frac{b}{2},  \label{eq:lambdas}
\end{equation}
with the associated ladder operators
\begin{eqnarray}
Z_{1} &=&x-y+i\left( p_{x}+p_{y}\right) ,  \nonumber \\
Z_{2} &=&x+y+i\left( p_{x}-p_{y}\right) ,  \nonumber \\
Z_{3} &=&Z_{1}^{\dagger },  \nonumber \\
Z_{4} &=&Z_{2}^{\dagger }.  \label{eq:Z's}
\end{eqnarray}
The only nonzero commutators are $\left[ Z_{1},Z_{4}\right] =\left[
Z_{2},Z_{3}\right] =4$. Every one of these ladder operators looks like the
sum of a creation operator for one vibrational mode and an annihilation
operator for the other which casts doubts on the existence of a square
integrable vacuum.

In the present case the adjoint matrix representations $\mathbf{H}$ and $%
\mathbf{H}_{1}$ of the operators $H_{d}$ and $H_{1}$, respectively, commute
in accordance to equation (\ref{eq:[mH,mA]=0}).

It is clear that the eigenvalues of $H$ are complex. Therefore, if $H\left|
\psi \right\rangle =E\left| \psi \right\rangle $ then it follows from $%
\left\langle \psi \right| H\left| \psi \right\rangle =\left\langle H\psi
\right| \left. \psi \right\rangle $ that $\left( E-E^{*}\right) \left\langle
\psi \right| \left. \psi \right\rangle =0$ and, consequently, $\left\langle
\psi \right| \left. \psi \right\rangle =0$. That is to say: the Hamiltonian
operator (\ref{eq:H_dim}) does not have square-integrable eigenfunctions
with the scalar product shown in equation (\ref{eq:Hermit_inner_prod}).

The function
\begin{equation}
\psi _{0}(x,y)=e^{-x^{2}/2+y^{2}/2},
\end{equation}
satisfies $Z_{1}\psi _{0}=Z_{2}\psi _{0}=0$ and $H\psi _{0}=\psi _{0}$.
Besides,
\begin{eqnarray}
HZ_{3}\psi _{0} &=&\left( 2-i\frac{b}{2}\right) Z_{3}\psi _{0},\;Z_{3}\psi
_{0}=2\left( x-y\right) \psi _{0},  \nonumber \\
HZ_{4}\psi _{0} &=&\left( 2+i\frac{b}{2}\right) Z_{4}\psi _{0},\;Z_{4}\psi
_{0}=2\left( x+y\right) \psi _{0},  \nonumber \\
HZ_{3}Z_{4}\psi _{0} &=&3Z_{3}Z_{4}\psi _{0},\;Z_{3}Z_{4}\psi
_{0}=Z_{4}Z_{3}\psi _{0}=4\left( x^{2}-y^{2}-1\right) \psi _{0},
\end{eqnarray}
in agreement with the general results of section~\ref{sec:algebraic}. A
curious result is that $Z_{1}Z_{3}^{n}\psi _{0}=0$ and $Z_{2}Z_{4}^{n}\psi
_{0}=0$ that follow from $\left[ Z_{1},Z_{3}\right] =0$ and $\left[
Z_{2},Z_{4}\right] =0$, respectively. In this case we obtain the eigenvalues
of Deguchi et al\cite{DFN19,DF19}:
\begin{eqnarray}
H\psi _{nm}^{(0)} &=&E_{nm}^{(0)}\psi _{nm}^{(0)},\;\psi
_{nm}^{(0)}=Z_{3}^{n}Z_{4}^{m}\psi _{0},\;  \nonumber \\
E_{nm}^{(0)} &=&n+m+1+(m-n)i\frac{b}{2},\;n,m=0,1,\ldots .
\end{eqnarray}

On the other hand, the function
\begin{equation}
\psi _{1}(x,y)=e^{x^{2}/2-y^{2}/2},
\end{equation}
satisfies $Z_{3}\psi _{1}=Z_{4}\psi _{1}=0$ and $H\psi _{1}=-\psi _{1}$. We
also have,
\begin{eqnarray}
HZ_{1}\psi _{1} &=&\left( -2-i\frac{b}{2}\right) Z_{1}\psi _{1},\;Z_{1}\psi
_{1}=2\left( x-y\right) \psi _{1},  \nonumber \\
HZ_{2}\psi _{1} &=&\left( -2+i\frac{b}{2}\right) Z_{2}\psi _{1},\;Z_{2}\psi
_{1}=2\left( x+y\right) \psi _{1},  \nonumber \\
HZ_{1}Z_{2}\psi _{1} &=&-3Z_{1}Z_{2}\psi _{1},\;Z_{1}Z_{2}\psi
_{1}=Z_{2}Z_{1}\psi _{1}=4\left( x^{2}-y^{2}+1\right) \psi _{1},
\end{eqnarray}
also in agreement with the general results of section~\ref{sec:algebraic}.
In this case we have $Z_{3}Z_{1}^{n}\psi _{1}=0$ and $Z_{4}Z_{2}^{n}\psi
_{1}=0$ for the same reason given above. The eigenvalues are quite different
from those shown above
\begin{eqnarray}
H\psi _{nm}^{(1)} &=&E_{nm}^{(1)}\psi _{nm}^{(1)},\;\psi
_{nm}^{(1)}=Z_{3}^{n}Z_{4}^{m}\psi _{1},\;  \nonumber \\
E_{nm}^{(1)} &=&-(n+m+1)+(m-n)i\frac{b}{2},\;n,m=0,1,\ldots .
\end{eqnarray}

Notice that the eigenvalues derived by means of the algebraic method and non
square-integrable eigenfunctions yield those obtained by Deguchi et al\cite
{DFN19,DF19} but not exactly those of Feshbach and Tikochinski\cite{FT77}.
Present algebraic method also reveals the difficulty in obtaining a suitable
vacuum state as argued by Bagarello et al\cite{BGR19,BGR20}. Not only the
eigenfunctions obtained are not square integrable but the operators $Z_{j}$
annihilate an infinite number of such functions.


\begin{thebibliography}{99}
\bibitem{B31}  H. Bateman, On dissipative systems and related variational
principles, Phys. Rev. 38 (1931) 815-819.

\bibitem{FT77}  H. Feshbach and Y. Tikochinsky, Quantization of the damped
harmonic oscillator, Transact. N.Y. Acad. Sci. 38, Ser. II (1977) 44-53.

\bibitem{DFN19}  S. Deguchi, Y. Fujiwara, and K. Nakano, Two quantization
approaches to the Bateman oscillator model, Ann. Phys. 403 (2019) 34-46.

\bibitem{DF19}  S. Deguchi and Y. Fujiwara, Square-integrable eigenfunctions
in quantizing the Bateman oscillator model, 2019.

\bibitem{BGR19}  F. Bagarello, F. Gargano, and F. Roccati, A no-go result
for the quantum damped harmonic oscillator, Ann. Phys. 383 (2019) 2836-2838.

\bibitem{BGR20}  F. Bagarello, F. Gargano, and F. Roccati, Some remarks on
few recent results on the damped quantum harmonic oscillator, Ann. Phys. 414
(2020) 168091.

\bibitem{FC96}  F. M. Fern\'{a}ndez and E. A. Castro, Algebraic Methods in
Quantum Chemistry and Physics, CRC, Boca Raton, New York, London, Tokyo,
(1996).

\bibitem{F15b}  F. M. Fern\'{a}ndez, Algebraic treatment of a simple model
for the electromagnetic self-force, 2015. arXiv:1509.00002 [quant-ph].

\bibitem{F15a}  F. M. Fern\'{a}ndez, Algebraic Treatment of PT -Symmetric
Coupled Oscillators, Int. J. Theor. Phys. 54 (2015) 3871-3876.
arXiv:1402.4473 [quant-ph].

\bibitem{F16c}  F. M. Fern\'{a}ndez, Algebraic treatment of non-Hermitian
quadratic Hamiltonians, 2016. arXiv:1605.01662v3 [quant-ph].

\bibitem{F16a}  F. M. Fern\'{a}ndez, Symmetric quadratic Hamiltonians with
pseudo-Hermitian matrix representation, Ann. Phys. 369 (2016) 168-176.
arXiv:1509.04267 [quant-ph].

\bibitem{F16b}  F. M. Fern\'{a}ndez, Algebraic treatment of the
Pais-Uhlenbeck oscillator and its PT-variant, 2016.

\bibitem{F19}  F. M. Fern\'{a}ndez, On the spectrum of an oscillator in a
magnetic field, 2019. arXiv:1909.04175 [quant-ph].

\bibitem{CGHS12}  E. Calliceti, S. Graffi, M. Hitrik, and J. Sj\"{o}strand,
Quadratic PT -symmetric operators with real spectrum and similarity to
self-adjoint operators, J. Phys. A 45 (2012) 444007.
\end{thebibliography}
\end{document}